\documentclass[%
reprint, twocolumn
superscriptaddress,
nofootinbib,
 amsmath,amssymb,
 aps,
]{revtex4-2}

\usepackage{graphicx}
\usepackage{bm}
\usepackage{hyperref}
\usepackage{xcolor}
\usepackage{amsmath}
\usepackage{comment}

\let\Re\relax \let\Im\relax
\DeclareMathOperator{\Re}{Re}
\DeclareMathOperator{\Im}{Im}
\DeclareMathOperator{\Tr}{Tr}

\DeclareMathOperator{\cn}{cn}

\def\be{\begin{equation}}
\def\ee{\end{equation}}
\def\ba{\begin{eqnarray}}
\def\ea{\end{eqnarray}}

\DeclareMathOperator{\atan}{atan}

\renewcommand{\H}{{\cal H}}
\newcommand{\PT}{{\cal PT}}

\begin{document}

\title{Instantons, analytic continuation, and $\mathcal{PT}$-symmetric field theory}
\author{Scott Lawrence}
\email{scott.lawrence-1@colorado.edu}
\affiliation{Department of Physics, University of Colorado, Boulder, CO 80309, USA}
\author{Christian Peterson}
\affiliation{Deptartment of Physics, University of Colorado at Colorado Springs, Colorado Springs, CO 80918, USA}
\author{Paul Romatschke}
\affiliation{Department of Physics, University of Colorado, Boulder, CO 80309, USA}
\affiliation{Center for Theory of Quantum Matter, University of Colorado, Boulder, Colorado 80309, USA}
\author{Ryan Weller}
\affiliation{Department of Physics, University of Colorado, Boulder, CO 80309, USA}

\begin{abstract}
    Ordinary Hermitian $\lambda \phi^4$ theory is known to exist in $d<4$ dimensions when $\lambda>0$. For negative values of the coupling, it has been suggested that a physical meaningful definition of the interacting theory can be given in terms of ${\cal PT}$-symmetric field theory. In this work, we critically re-examine the relation between analytically continued Hermitian field theory with quartic interaction, and ${\cal PT}$-symmetric field theory, including $O(N)$ models. We find that in general ${\cal PT}$-symmetric field theory \textit{does not} correspond to the analytic continuation of the Hermitian theory, except at high temperature where the instanton contribution present in the analytically continued theory can be neglected.
\end{abstract}
\maketitle

\section{Introduction}

Hermitian field theory is built around the presence of a Hermitian Hamiltonian that is bounded from below. In quantum mechanics, it has long been known that Hermiticity and a lower-bounded potential are sufficient to guarantee a real and lower-bounded spectrum of the Hamiltonian, thus providing the basis for modern quantum field theory. However, it has been found that somewhat weaker conditions than Hermiticity and boundedness, namely symmetry under parity $\cal P$ and time reversal $\cal T$, still result in real and semi-definite energy eigenspectra \cite{Bender:1998ke}. In fact, it has been proved that $\cal PT$-symmetry is sufficient to guarantee real spectra in quantum mechanics \cite{Dorey:2001uw}, showing that Hermiticity is not a necessary condition.

A natural generalization of $\cal PT$-symmetric quantum mechanics is $\cal PT$-symmetric quantum field theory, which is a fairly recent area of study. In a series of articles, it has been suggested that Hermitian field theory with a quartic interaction and negative coupling constant can be related to ${\cal PT}$-symmetric field theory \cite{Bender:2021fxa,Mavromatos:2021hpe,Grunwald:2022kts,Ai:2022csx}. In particular, in \cite{Ai:2022csx} it is conjectured that the partition function $Z_{\cal H}$ of the Hermitian field theory can be related to the partition function of the ${\cal PT}$-symmetric field theory $Z_{\cal PT}$ in $d>0$ dimensions via
\begin{equation}
\label{ABS}
    \ln Z_\PT(g)={\rm Re}\ln Z_\H(\lambda=-g)\,,
\end{equation}
where $\lambda$ is the coupling constant of the Hermitian theory with $\lambda\phi^4$ interaction, and $Z_\H$ refers to the analytic continuation of the Hermitian theory's partition function.

If the ABS conjecture (\ref{ABS}) holds for quantum field theory with quartic interaction in general dimensions $d$, this would provide meaning for quantum field theories in situations where the potential becomes unbounded, in particular scalar quantum field theory in four dimensions, see e.g.~Refs.~\cite{Romatschke:2022jqg,Romatschke:2022llf,Grable:2023paf}. For this reason, it is interesting to study the precise relation between analytically continued Hermitian and $\cal PT$-symmetric field theory. In particular, we aim to study the ABS conjecture (\ref{ABS}) in cases where both sides of the equation can be evaluated. This is particularly easy in $d=0$, where Ref.~\cite{Ai:2022csx} already noted that the partition functions fulfill the relation
\begin{equation}
    \label{conj}
    Z_{\cal PT}(g)=\Re Z_\H(\lambda=-g)\text{,}
\end{equation}
instead of (\ref{ABS}).

In this work we examine these two conjectures in the $d=1$ case; that is, quantum mechanics. Here high-precision numerical calculations are possible, and we find that neither conjecture holds at all values of the dimensionless parameter $\beta^3 g$. However, at high temperatures (equivalently at weak coupling), the second conjecture (\ref{conj}) holds to high precision. We provide numerical evidence and, by considering the semiclassical expansion in $\beta^3 g$, an argument from complex analysis indicating that the failure of that conjecture to hold at low temperatures (strong coupling) is due to the presence of nonperturbative bounce\footnote{We will be dealing with periodic instanton solutions that in the literature are referred to as "bounces", hence we will use the term bounce in the following.} contributions to the analytically continued partition function.

The remainder of this paper is structured as follows. We confirm the result (\ref{conj}) for $d=0$ and consider the extension to multi-component scalar fields (also known as the O(N) model) in Section~\ref{sec:warmup}. We then continue in Section~\ref{sec:numerical} to study the quantum mechanical ($d=1$) case, where the partition function for both sides of (\ref{ABS}) can be obtained numerically to high precision. We show that there is no correspondence of the form of (\ref{ABS}); however, the analog of (\ref{conj}) is true to high precision at low temperatures. The numerical evidence indicates that the difference between the two sides of (\ref{conj}) is due to an extra nonperturbative bounce contribution in the analytically continued partition function. Working in the path integral formalism, we provide an explanation for this fact in Section~\ref{sec:pi}. Finally, we discuss the implications of our findings in Section~\ref{sec:disc}.

\section{The one-site model}
\label{sec:warmup}

As a warm-up to quantum field theory, let us first discuss the limiting case of zero dimensions. This section will focus on complex-analytic arguments to reveal the behavior of the partition function without the need to find closed-form expressions; explicit calculations are provided in appendix~\ref{app}.

\subsection{Warm-up: One component}
\begin{figure*}
    \centering
    \includegraphics[width=0.95\linewidth]{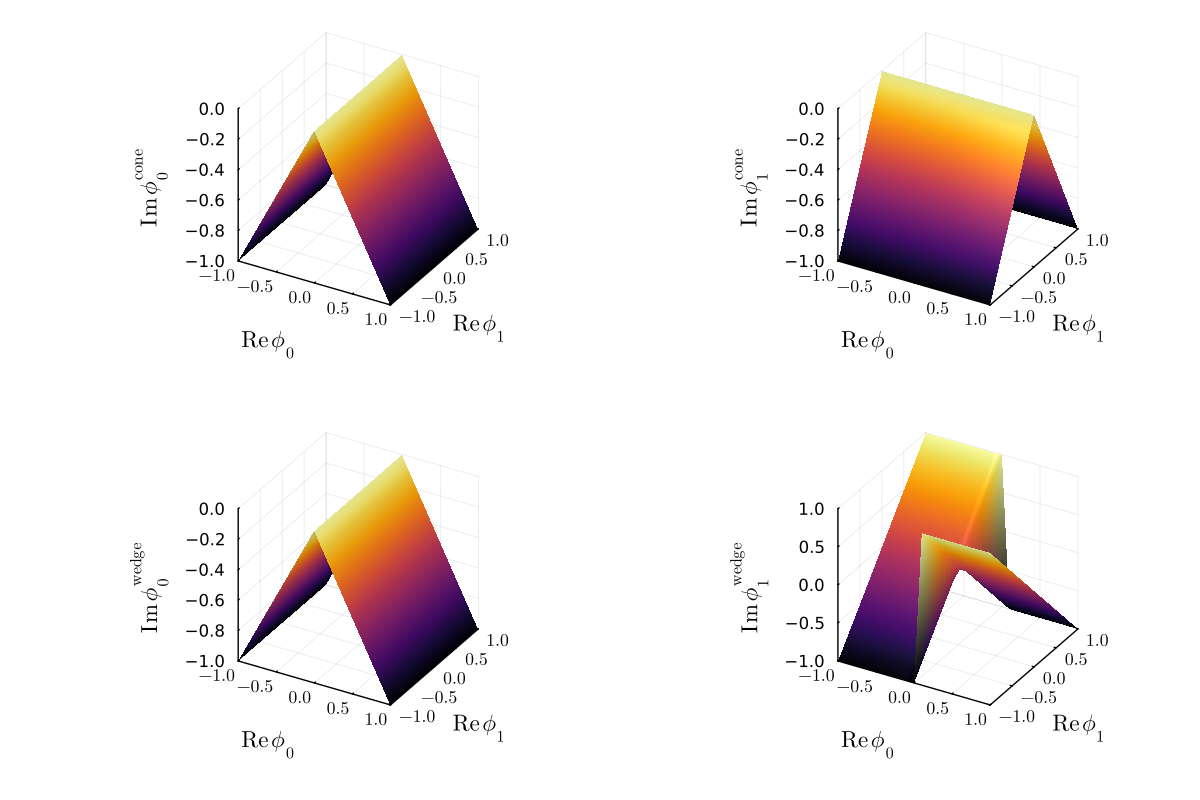}
    \caption{The `cone' and `wedge' contours considered throughout this paper, in the case of a two-component scalar field. The cone contour is shown on the top by plotting $\Im \phi_1(\Re \phi_1,\Re \phi_2)$ and $\Im \phi_2(\Re \phi_1,\Re \phi_2)$. The wedge contour is similarly portrayed on the bottom.}
    \label{fig:contours}
\end{figure*}

The partition function for standard Hermitian field theory in $d=0$ becomes a single integral over the field,
\begin{equation}\label{eq:Z0_exact}
Z^{d=0}(\lambda)=\int_{-\infty}^\infty \frac{d\phi}{\sqrt{2 \pi}} e^{-\lambda \phi^4}=\frac{2 \Gamma\left(\frac{5}{4}\right)}{\left(4 \pi^2 \lambda\right)^{\frac{1}{4}}}\text.
\end{equation}
As written this partition function is only defined for $\Re \lambda > 0$; however, it may be extended to all $\lambda \ne 0$ (although not uniquely, due to a branch point at the origin) by analytic continuation. This is clearly seen from the right-hand side of (\ref{eq:Z0_exact}); however, without access to a closed-form solution for the integral, the analytic continuation is still easily accomplished by deforming the contour of integration to preserve the convergence of the integral as $\lambda$ is rotated from $\mathbb R_+$ to elsewhere on the complex plane. In a slight abuse of notation, for $\lambda = \Lambda e^{i \theta}$, we may write
\begin{equation}
Z^{d=0}\big(\lambda = \Lambda e^{i \theta}\big)=\int_{-\infty e^{i \theta / 4}}^{\infty e^{i \theta/4}} \frac{d\phi}{\sqrt{2 \pi}} e^{-\lambda \phi^4}\text.
\end{equation}
This expression also makes clear the non-uniqueness of the analytic continuation. For example, for negative real values of $\lambda$, the analytic continuation may involve integrating either along a contour for which $\phi$ is proportional to $e^{i \pi / 4}$, or one proportional to $e^{-i \pi / 4}$. However, the integrals along these two contours are related by complex conjugation: the real parts do not differ.

So much for the analytic continuation of the partition function; now we consider the ``\PT-symmetric'' version. This version of the partition function is intended to be real, and to correspond to the case $\lambda < 0$, but here the integral no longer converges. To obtain a well defined partition function, we will deform the domain of integration from the real line to some other contour $\gamma$. In general, a contour will yield a convergent integral at $\lambda < 0$ if $\phi^4$ approaches $-\infty$ in either direction along the contour. From Cauchy's integral theorem, two such contours will yield the same partition function if one can be smoothly deformed into the other without passing through any regions where $e^{g \phi^4}$ diverges.

We can satisfy all these constraints by defining the $\PT$-symmetric theory as
\be
Z^{d=0}_{\cal PT}(g)=\int_{\gamma_\PT}\frac{d\phi}{\sqrt{2 \pi}} e^{g \phi^4}\text,
\ee
with a contour $\gamma_\PT$ defined by\be
\label{stokes}
\phi(s)=\begin{cases}
    s e^{i \frac\pi 4} & s < 0\\
    s e^{-i \frac \pi 4} & s \ge 0
\end{cases}
\ee
with $s\in \mathbb{R}$ parametrizing the contour.

To relate the Hermitian and $\PT$-symmetric partition functions in this $d=0$, one-component case, it is helpful to define four `partial' integration contours, each connecting the origin to some asymptotic region where $\phi^4 \rightarrow -\infty$. Each contour is parameterized by $s \in [0,\infty)$:
\begin{eqnarray}
    \gamma_1 : \phi(s) &=& s e^{i \frac \pi 4}\\
    -\gamma_2 : \phi(s) &=& s e^{i \frac {3\pi} 4}\\
    -\gamma_3 : \phi(s) &=& s e^{i \frac {5\pi} 4}\\
    \gamma_4 : \phi(s) &=& s e^{i \frac {7\pi} 4}
\end{eqnarray}
These four contours each lie in a different quadrant of the complex plane, and are numbered accordingly. Finally note that $\gamma_2$ and $\gamma_3$ have reversed orientation, so that the integration is taken from complex infinity to the origin, rather than \emph{vice versa}. As a result, each contour is oriented so that integration is performed from ``right to left'' on the complex plane.

With these definitions, the contour defining the $\PT$-symmetric theory above is given by $\gamma_\PT = \gamma_3 + \gamma_4$. The (clockwise) analytic continuation is defined by integrating instead along $\gamma_{\mathrm{ac}} = \gamma_3 + \gamma_1$. Denoting for brevity $I_k = \int_{\gamma_k} e^{\phi^4}$, we see that the various partial integrals are related by
\begin{equation}
I_1 = I_2^* = I_3 = I_4^*
\text.
\end{equation}
A short calculation therefore relates the (analytically continued) Hermitian and $\PT$-symmetric partition functions in this case: the $\PT$-symmetric partition function $Z=\frac{ \Gamma\left(\frac{5}{4}\right)}{\left(\pi^2 g\right)^{\frac{1}{4}}}$ is simply given by the real part of the analytically continued Hermitian partition function:
\be
\label{d0relation}
Z^{d=0}_{\cal PT}(g)={\rm Re} \left(Z^{d=0}(\lambda=-g)\right)\text.
\ee
As noted in~\cite{Ai:2022csx}, this relation is \emph{different} from the conjecture (\ref{ABS}), which involves the logarithm of the partition function.

\subsection{N-component scalars}

We may now investigate the relation between Hermitian and ${\mathcal PT}$-symmetric field theory for d=0 for N-component scalars $\vec{\phi}=\left(\phi_1,\phi_2,\ldots\phi_N\right)$. In this case, the partition function for the Hermitian field theory is defined as
\be
\label{d0hermitz}
Z_{N}^{d=0}=\int \frac{d\vec{\phi}}{(2\pi)^{\frac{N}{2}}} e^{-\frac{\lambda}{N} (\vec{\phi}^2)^2}\text.
\ee

The partition function for ${\cal PT}$-symmetric QFT is defined by
\be
Z^{d=0}_{{\cal PT},N}=\int \frac{d\vec{\phi}}{(2\pi)^{\frac{N}{2}}} e^{\frac{g}{N}\left(\vec{\phi}^2\right)^2}\,,
\ee
where the integration is \textit{not} on the real axis, but in the complex plane. For pedagogical reasons, it is useful to first consider the explict case of N=2 (two component scalar fields) where $\vec{\phi}=(\phi_0,\phi_1)$. The $\cal PT$-symmetric field theory is then defined by using the parametrization (\ref{stokes}) for both $\phi_0,\phi_1$, effectively parametrizing a 'cone' in the complex 4-dimensional parameter space (see Figure~\ref{fig:contours}). Explicitly, one has
\begin{eqnarray}
\label{cone}
\phi_0&=&s \left(e^{\frac{i\pi}{4}}\theta(-s)+e^{-\frac{i\pi}{4}}\theta(s)\right)\text{, and}\nonumber\\
\phi_1&=&t \left(e^{\frac{i\pi}{4}}\theta(-t)+e^{-\frac{i\pi}{4}}\theta(t)\right)\,,
\end{eqnarray}
with $s,t\in \mathbb{R}$. The resulting $\cal PT$-symmetric path integral for N=2 therefore is
\be
Z^{d=0}_{{\cal PT},N=2} = \int_0^\infty \frac{ds dt} {(2\pi)} \left[e^{-\frac{g}{2}\left(s^2+t^2\right)^2}\cos\left(\frac{\pi}{2}\right)+e^{-\frac{g}{2}\left(s^2-t^2\right)^2}\right]\,.
\ee
It is straightforward to see that the $N=2$ $\cal PT$-symmetric partition function diverges, because there is a flat direction $s=t$ in the integrand along with the action is constant. In fact this finding generalizes to any integer $N>1$ when fields are quantized on the cone as a repeated application of (\ref{cone}). As a consequence, we find that for $N>1$, the $\cal PT$-symmetric partition function obeys neither the ABS conjecture (\ref{ABS}) nor the relation (\ref{conj}) proved for N=1.

However, it is possible to give a meaningful definition of the path integral with negative coupling constant for the case $N>1$. To this end, consider again the case of N=2, but now parametrize fields on a 'wedge' in the complex 4-dimensional parameter space (see again Figure~\ref{fig:contours}). Explicitly, one then has the `wedge' contour defined by
\begin{eqnarray}
\phi_0&=&s \left(e^{\frac{i\pi}{4}}\theta(-s)+e^{-\frac{i\pi}{4}}\theta(s)\right)\text{, and }\nonumber\\
\phi_1&=&t \left(e^{\frac{i\pi}{4}}\theta(-s)+e^{-\frac{i\pi}{4}}\theta(s)\right)\text,\label{eq:wedge}
\end{eqnarray}
where again $s,t\in \mathbb{R}$.

As in the one-site case, we will now show that this contour arises as the real part of the analytic continuation to negative $\lambda$ of the Hermitian theory. Let us introduce some notation to make working with simple multi-dimensional integration contours tractable. Given two one-dimensional contours $\gamma_a$ and $\gamma_b$, denote by $\gamma_a \times \gamma_b$ the two dimensional contour consisting of points $(z_a, z_b) \in \mathbb C^2$ with $z_a \in \gamma_a$ and $z_b \in \gamma_b$. In this notation, the wedge contour (\ref{eq:wedge}) defined above may be written
\begin{equation}
    \gamma_{\mathrm{wedge}} = \gamma_3 \times (\gamma_3 + \gamma_1) + \gamma_4 \times (\gamma_2 + \gamma_4).
\end{equation}
Meanwhile, the clockwise analytic continuation yields a different contour:
\begin{equation}
\gamma_{\mathrm{ac}} = (\gamma_2 + \gamma_4) \times (\gamma_2 + \gamma_4)
\text.
\end{equation}
For brevity, denote the integral along the contour $\gamma_i \times \gamma_j$ by $I_{ij}$. The clockwise analytic continuation is equal to $I_{22} + I_{24} + I_{42} + I_{44}$, while the integral along the wedge contour is
\begin{equation}
    \int_{\gamma_{\mathrm{wedge}}} \frac 1 {2\pi} e^{-S} 
    =
    I_{33} + I_{31} + I_{42} + I_{44}\text.
\end{equation}
As before, to relate these two note that we have the following relations amongst the various partial integrals:
\begin{eqnarray}
&&I_{11} = I_{22}^* = I_{33} = I_{44}^*\text{, and}\\
&&I_{13} = I_{24}^*
\text.
\end{eqnarray}
From this it follows that \be\Re \left(I_{22} + I_{23} + I_{32} + I_{33}\right) = I_{33} + I_{31} + I_{42} + I_{44}\text,\ee confirming the desired identity. The same proof holds without modification for the case of three or more components; all new components are treated as $\phi_1$.

To review: in the $N$-component, $d=0$ case, we have examined two contours on which we could attempt to define the partition function. The `cone' contour---arguably the more obvious generalization of the $N=1$ case---results in an undefined partition function. The `wedge' contour corresponds exactly to the real part of the analytic continuation of the original, Hermitian theory to negative couplings. This establishes an analog of (\ref{conj}) for multi-component theories in $0$ dimensions.

Finally, a brief note on the ABS conjecture itself. Because the analytic continuation of $\log Z^{d=0}$ has a non-zero imaginary part, (\ref{conj}) implies that the ABS conjecture (\ref{ABS}) does not hold at any finite $N$. However, in the large-$N$ limit, both the real and imaginary parts of the free energy---for both the analytically continued and the $\PT$-symmetric theories---may be expanded in powers of $N$. The leading terms in the real parts are proportional to $N$, but because of the logarithm the leading term in the imaginary part can at most be $O(N^0)$. As a result, the conjecture (\ref{conj}) directly implies the ABS conjecture in the large-$N$ limit.

Thus at large $N$ and large $N$ only, the Hermitian and wedge-contour parametrized partition functions for the $d=0$ case are related through the ABS conjecture (\ref{ABS}), as well as the modified conjecture (\ref{conj}).

\section{Numerical comparison}\label{sec:numerical}

\begin{figure*}
    \centering
    \includegraphics[width=0.48\linewidth]{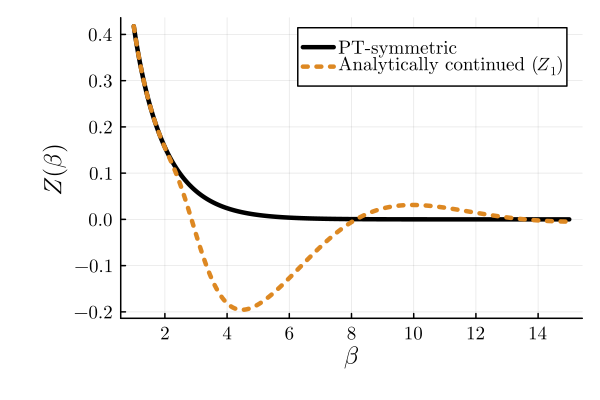}
    \hfil
    \includegraphics[width=0.48\linewidth]{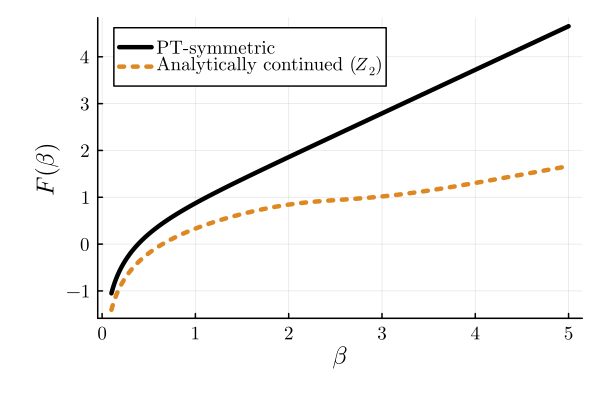}
    \caption{Numerical checks of the two conjectures. The conjecture (\ref{conj}) is examined in the left panel, where very precise agreement in the partition functions at small $\beta$ gives way to unphysical behavior in the analytically continued theory at low temperatures. The ABS conjecture itself is checked in the right panel, where the logarithm of the partition functions is plotted.}
    \label{fig:compare}
\end{figure*}

Let us now discuss the case of a single-component quantum field with a quartic interaction in $0+1$ dimensions, with both positive coupling sign (the Hermitian theory) and negative coupling sign (the $\cal PT$-symmetric theory). First we will define the different theories under consideration---two theories constructed via analytic continuation, and the $\cal PT$-symmetric theory. Then we detail numerical schemes for computing a high-precision partition function in all three cases, and finally we perform a comparison, the results of which indicate that neither construction via analytic continuation is equivalent to the $\cal PT$-symmetric theory. One, however, is sufficiently closely related to merit further examination; this is done in the subsequent section.

The Hermitian theory is defined from the Hamiltonian
\begin{equation}\label{eq:H_H}
H_\H = p^2 + \frac \lambda 4 x^4\text,
\end{equation}
from which a partition function $Z_\H(\beta; \lambda) \equiv \Tr e^{-\beta H_\H(\lambda)}$ is obtained. As written, this function is defined only on the right half-plane of complex $\lambda$; elsewhere the Hamiltonian is unbounded below and the trace diverges. However, it follows from dimensional analysis that the partition function depends only on the combination $\beta^3 \lambda$. As a result we find that $Z_{\cal H}(\beta, \lambda) = Z_{\cal H}(\beta\lambda^{1/3}, 1)$. We can use this relation to analytically continue $Z_{\cal H}$ to values of $\lambda$ in the left half-plane.

The analytically continued Hermitian partition function has a branch point at $\lambda = 0$, and as a result the analytic continuation is not unique. Following the conjecture, we analytically continue to negative values of $\lambda$ along both clockwise and counterclockwise paths. The two resulting partition functions may be computed as
\begin{eqnarray}
Z_{\mathrm{cw}}(\beta,g) &\equiv& \Tr e^{-\beta e^{i \frac{\pi}{3} }H_{\cal H}(g)}
    \text{, and }\\
Z_{\mathrm{ccw}}(\beta,g) &\equiv&  \Tr e^{-\beta e^{-i \frac{\pi}{3} }H_{\cal H}(g)}
\text,
\end{eqnarray}
where as in the previous section we have defined $g=-\lambda$ to be the wrong-sign coupling.

From these analytically continued partition functions, we can define a candidate ${\cal PT}$-symmetric theory either by averaging either the two partition functions, or their logarithms. The former yields a partition function analogous to the one constructed in the $d=0$ case (\ref{conj}):
\begin{equation}\label{eq:Z_1}
    Z_{\mathrm{1}} \equiv \Re \Tr e^{-\beta e^{i \frac{\pi}{3} }H_{\cal H}(g)}
    \text.
\end{equation}
The latter approach yields the partition function of the ABS conjecture (\ref{ABS}):
\begin{equation}\label{eq:Z_2}
    Z_{\mathrm{2}} \equiv  \left|\Tr e^{-\beta e^{i \frac{\pi}{3} }H_{\cal H}(g)}\right|
    \text.
\end{equation}

The $\PT$-symmetric theory is defined by quantizing the Hamiltonian~(\ref{eq:H_H}) at negative coupling $\lambda = -g$, on a contour other than the real line. We will parameterize the contour $x(s) \in \mathbb C$ by some $s \in \mathbb R$. A wide variety of contours yield the same spectrum; it is sufficient to consider any smooth contour $x(s)$ with $|x| = |s|$ and obeying
\begin{equation}
\exp \left(i \arg \lim_{s \rightarrow \pm\infty} x(s)\right)
=-i e^{\pm i \frac\pi 4}
\text.
\end{equation}
A common choice is to take the contour to be the sum of two linear pieces going through the origin:
\begin{equation}
x(s) = \begin{cases}
s e^{-i \frac \pi 4} & s \ge 0\\
s e^{i \frac \pi 4} & s < 0
\end{cases}
    \text.
\end{equation}
From the spectrum of the Hamiltonian (\ref{eq:H_H}), the partition function of the $\PT$-symmetric theory is obtained in the usual way:
\begin{equation}\label{eq:Z_PT}
    Z_\PT = \sum_n e^{- \beta E_n}\text.
\end{equation}

All three theories defined above are amenable to high-precision numerical calculation. In the case of the first two, we determine the eigenenergies of $H_{\cal H}$ by expressing that Hamiltonian in the occupation number basis of the harmonic oscillator, and numerically diagonalizing. A truncation of the first $100$ states of the harmonic oscillator is found to yield eigenvalues of sufficient precision for this study; all plots and numerical results reported herein come from a truncation of the first $10^3$ states. With these eigenenergies determined, it is straightforward to evaluate either (\ref{eq:Z_1}) or (\ref{eq:Z_2}) numerically; the sums exhibit exponential convergence even at negative coupling.

In order to obtain (\ref{eq:Z_PT}), we exploit the exact duality demonstrated in~\cite{Jones:2006qs}: the spectrum of the $\PT$-symmetric Hamiltonian, quantized on a suitable contour, is equal to that of
\begin{equation}
    H_{\mathrm{dual}} = p^2 - x + 4 x^4 \text.
\end{equation}
The spectrum of $H_{\mathrm{dual}}$ is obtained, as before, by diagonalizing the Hamiltonian expressed in the occupation number basis of the harmonic oscillator. As before, $100$ states are sufficient for this study, and $10^3$ are used for all results hereafter.

We are now prepared to compute the three different partition functions and compare. The results of this evaluation are shown in Figure~\ref{fig:compare}. The left panel is a check of the conjecture (\ref{conj}), which is clearly seen to fail at large $\beta$ where the analytically continued partition function becomes unphysically negative. The right panel checks the ABS conjecture (\ref{ABS}), where both partition functions exhibit physical behavior, but do not match.

\begin{figure}
    \centering
    \includegraphics[width=0.95\linewidth]{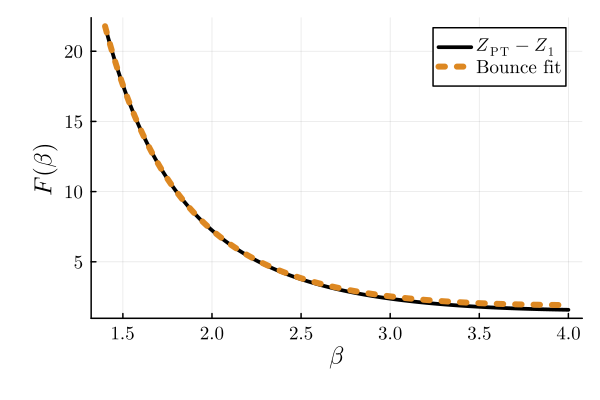}
    \caption{Detailed study of the failure of (\ref{conj}). The logarithm of the difference between the two partition functions is plotted at small $\beta$ (equivalent to small coupling $\lambda$). A fit to the functional form (\ref{eq:fit}) is performed over the range $\beta \in [1.4,1.8]$, and compared with the numerical form over $\beta \in [1.4,4.0]$. The fit agrees to high precision and generalizes well to those smaller temperatures.}
    \label{fig:fit}
\end{figure}

Although the left panel refutes (\ref{conj}), there is still surprising and suggestive agreement at small $\beta$ (high temperatures or equivalently, weak couplings). The precise agreement at small $\beta$ followed by sudden onset of disagreement is suggestive of nonanalytic behavior akin to that of $f(x) = e^{-1/x^2}$ near the origin. The logarithm of the difference between the two partition functions at small $\beta$ is shown in Figure~\ref{fig:fit}. To high precision, and across several orders of magnitude of the partition functions, this difference is found to be fit by a function
\begin{equation}\label{eq:fit}
    -\log \left(Z_\PT - Z_1\right) \approx
    \frac{p_1^3}{\beta^3} +
    \log (p_2 + p_3 \beta^3)
\end{equation}
with parameters $p_1 \approx 3.963$, $p_2 \approx 0.307$, and $p_3 \approx 0.035$.

The numerically observed form of the failure of (\ref{conj}) at small $\beta$ provides a clue as to the origin of the difference for high values of $\beta$, since it has the same parametric dependence on the coupling---of the form $e^{-1 / (\beta^3 \lambda)}$---as a bounce contribution~\cite{coleman1979uses}. The next section explores this further.

\section{Path integrals}\label{sec:pi}
To explain the relation between the ${\cal PT}$-symmetric theory and the analytically continued partition function $Z_1$, we switch from the Hamiltonian to the action formalism. The action of either theory is
\begin{equation}
    S = \int d\tau\,\left[\frac 1 2 \left(\frac{d\phi}{d\tau}\right)^2 + \lambda \phi^4\right]
    \text,
\end{equation}
although making sense of this in either the $\PT$-symmetric or analytically continued cases (where $\lambda < 0$) requires taking the path integral over an appropriate contour\footnote{Note that this is not the same as the process of ``quantizing on a contour'' that was used to define the $\PT$-symmetric Hamiltonian theory. For example, in the path integral formulation, $\phi(t)$ and $\phi(t')$ may live on two different contours in $\mathbb C$.}. We will see that the choice of contour is what makes the difference observed in the previous section: one contour corresponds to the analytically continued theory, and a different contour to the ${\cal PT}$-symmetric theory.

The first subsection below traces through the derivation of the path integral starting from the Hamiltonian formalism, showing that if the starting point is the analytically continued theory $Z_1$, one contour (analogous to the `wedge' discussed in Section~\ref{sec:warmup} above) is obtained, but if the starting point is the ${\cal PT}$-symmetric theory, the path integral must be performed over a different contour (the `cone'). Next we examine a lattice discretization of the path integral, and show numerically that it yields qualitatively similar results to the above. After reviewing some basic facts about Lefschetz thimbles and their intersection numbers, we show that the two contours have the same contribution from the trivial saddle point at the origin, and therefore must differ in their contribution from some other saddle point. The final subsection examines the saddle points of the action and performs a semiclassical expansion around the nontrivial ones; we find that this expansion matches the functional form (and, to decent precision, the exponent) found by fitting the difference of partition functions above.

\subsection{Two contours}

First let us perform a loose derivation of a path integral for the $\PT$-symmetric theory, beginning with the Hamiltonian $H_\PT = H_\H(\lambda = -g)$. The derivation proceeds in the usual way, but we use the following resolution of the identity:
\begin{equation}
    1 = \int_{\gamma_\PT} dx\,|x\rangle\langle x|
    \text.
\end{equation}
As a result, the partition function $Z_\PT \equiv \Tr e^{-\beta H_\PT}$ reads
\begin{equation}\label{eq:Z_PT_int}
    Z_\PT = \int_{x(t) \in \gamma_\PT} \mathcal D x(t) \, e^{-S(x;\lambda=-g)}
\end{equation}
where at every time $t$, the position $x(t)$ is required to be valued not on the real line, but on the deformed contour $\gamma_\PT$ used to quantize the $\PT$-symmetric theory.

In the $d=0$ case, we were able to show in Section~\ref{sec:warmup} that an analogous integral corresponded to the real part of the analytic continuation of the original partition function, but critically, this held only for $N=1$. For a multi-component field, obtaining the real part of the analytic continuation requires the use of the wedge contour, defined by (\ref{eq:wedge}) and depicted in Figure~\ref{fig:contours}. The same derivation holds here without modification.

\subsection{On the lattice}

For Hermitian theories, the partition function in quantum mechanics can also be defined as a path integral over real values of the field $\phi$,
\be
Z^{d=1}_H(\lambda)=\int {\cal D}\phi \, e^{-S}\text.
\ee The path integral may be discretized  by dividing the imaginary time interval into K sites \cite{Laine:2016hma}
\be
Z^{d=1}(\lambda)=\lim_{K\rightarrow \infty}\int \prod_{i=1}^K \frac{d\phi_i}{\sqrt{2\pi \varepsilon}}e^{-S_{\mathrm{lat}}}\text,
\ee
where the lattice action is defined as
\be
S_{\mathrm{lat}}=\varepsilon\sum_{i=1}^K \left[\frac{(\phi_{i}-\phi_{i+1})^2}{2\varepsilon^2}+\lambda \phi_i^4\right]\text,
\ee
with $\varepsilon=\frac{\beta}{K}$ and periodic boundary conditions $\phi_{K+1}=\phi_1$. In this form, the partition function is amenable to numerical computation for given values of $\lambda,\beta$. The number of sites $K$ must be chosen such that $\varepsilon\ll 1$ in order to be close to the continuum limit of the theory. 
In practice, we find that in units where $\lambda=1$, the choice $\varepsilon<0.5$ gives acceptable quantitative results.

\begin{figure}
    \centering
    \includegraphics[width=0.95\linewidth]{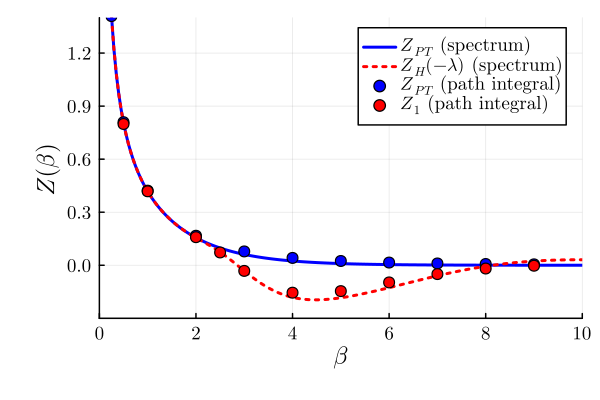}
    \caption{Same as figure \ref{fig:compare}, but comparing results for $Z^{d=1}$ from the Hamiltonian spectrum to those from the path integral. See text for details.}
    \label{fig:two}
\end{figure}

For the $\cal PT$-symmetric theory, the integration domain is not real. As with the case of d=0, one can, however, choose each $\phi_i$ to be given by (\ref{stokes}), such that with $\chi=-\frac{\pi}{4}$
\be
\label{ZPTd1def}
Z^{d=1}_{\cal PT}(g)=\int \prod_{i=1}^K \frac{d \phi_i}{\sqrt{2\pi \varepsilon}}e^{-S_{\cal PT}}\,,
\ee
with the $\PT$-symmetric form of the lattice action $S_\PT(g) = S_{\mathrm{lat}}(\lambda = -g)$.
This is analogous to the `cone' contour of previous sections, except now defined for multiple sites rather than multiple components of the field. The resulting path integral is convergent, but somewhat unwieldy to implement. Note that it instead of (\ref{stokes}) is possible to choose complex integration contours without kinks such that the resulting path integral can be cast in form of a real-integration domain with a real action \cite{Bender:2006wt}, which is numerically preferable to (\ref{ZPTd1def}). However, we find that (\ref{ZPTd1def}) with just four sites ($K=4$) gives qualitatively acceptable results for $Z^{d=1}_{\cal PT}(g)$ for $\beta<3$, see figure \ref{fig:two}.

Recalling the discussion in Section~\ref{sec:warmup}, it was found that in the case of d=0, the `cone' contour of integration did not reproduce the analytically continued Hermitian theory for more than one field $N>1$. However, it was found that a `wedge' contour (\ref{eq:wedge}) faithfully gave the correct analytic continuation. For this reason, we consider a \textit{different} path integral given by choosing the fields to lie on the wedge (\ref{eq:wedge}) where instead of $i=0,1,2,\ldots,N$ the index in (\ref{eq:wedge}) now refers to the site location, e.g. $i=0,1,2,\ldots,K$. With $\chi=-\frac{\pi}{4}$, one finds 
\be
\label{Zd1wedge}
Z^{d=1}_{\rm wedge}(g)=\int_0^\infty \frac{ds_0}{\sqrt{2\pi\varepsilon}} 
\int_{-\infty}^\infty \prod_{i=1}^{K-1} \frac{ds_i}{\sqrt{2\pi \varepsilon}}\left(e^{-S^+}+e^{-S^-}\right)
\,,\quad
\ee
\be
S^\pm =\pm\frac{i \pi K}{4}+\varepsilon\sum_{i=0}^{K-1} \left[\pm i\frac{(s_{i}-s_{i+1})^2}{2\varepsilon^2}+g s_i^4\right]\,.
\ee
The wedge path-integral is convergent, and can be evaluated numerically using efficient numerical integrators such as VEGAS \cite{Lepage:2020tgj} on modern CPUs for $K\lesssim 20$. Results for $K=10$ for $Z_{\rm wedge}(g)$ are shown in figure \ref{fig:two}, suggesting that the wedge path-integral indeed corresponds to the analytic continuation of the Hermitian theory to negative coupling.

\subsection{Intersection numbers}

To understand the origin of the difference between the integrals on the two contours, we must first review the properties of Lefschetz thimbles (see~\cite{Witten:2010cx} for a more detailed exposition). We will assume that the model has been defined on a finite number of degrees of freedom, as in the lattice models of the previous section.

Given a holomorphic action $S(z)$ of fields $z \in \mathbb C^N$, we define the \emph{upward flow} according to
\begin{equation}\label{eq:hgf}
    \frac{d z_i}{d t} = \left(\frac{\partial S}{\partial z_i}\right)^*
    \text.
\end{equation}
The upward flow has the important property that along it, the imaginary part of the action is constant, while the real part of the action monotonically increases.

The flow vanishes only at solutions to the classical equations of motion---i.e., the saddle points. To each saddle point $z^{(\sigma)}$ is associated a \emph{Lefschetz thimble} $\mathcal J_\sigma$: a $K$-dimensional manifold consisting of the union of all solutions $z(t)$ to (\ref{eq:hgf}) obeying $\lim_{t \rightarrow -\infty} z(t) = z^{(\sigma)}$. We may similarly define an \emph{anti-thimble} as the union of all solutions obeying $\lim_{t \rightarrow +\infty} z(t) = z^{(\sigma)}$. Note that the integral of $e^{-S}$ along a thimble is finite, while the integral along the anti-thimble diverges.

Any integration contour that begins and ends at complex infinity is homologous to some linear combination of thimbles, with integer coefficients. In particular, this implies that the integral of any holomorphic function (including of course $e^{-S}$) along that contour is the sum of the integrals taken along those thimbles:
\begin{equation}\label{eq:thimble-decomposition}
    \int_\gamma f(z) = \sum_\sigma n_\sigma \int_{\mathcal J_\sigma} f(z)
\end{equation}
Once an integration contour has been expressed as a linear combination of thimbles, we may perform a saddle point approximation on each thimble. The contribution of the thimble $\mathcal J_\sigma$ will be proportional (up to a Jacobian factor) to $e^{-S(z^{(\sigma)})}$.

The integers $n_\sigma$ in (\ref{eq:thimble-decomposition}) are termed \emph{intersection numbers}. In practice we commonly find $n_\sigma \in \{0, \pm 1\}$. If the integrals on the cone and wedge contour are to differ, then those two contours must have differing intersection numbers. If the difference between those two contours is to be surpressed by a factor of $e^{-1 / \lambda \beta^3}$, then the difference in intersection numbers must not be at the origin, but at a sub-dominant saddle point.

In principle, the intersection numbers may be obtained by evolving the integration contour of interest according to (\ref{eq:hgf}). In the limit of long-flow times, this will approach a fixed-point manifold exactly equal to some integer combination of the various Lefschetz thimbles. This is typically not a practical procedure, but it provides a useful trick for establishing that an intersection number is $0$, as follows. Recall that the upward flow only increases the real part of the action. If, for every point $z$ in the integration contour of interest, the real part of the action is already larger than that at a saddle point $z^{(\sigma)}$, then the associated intersection number is necessarily $n_\sigma = 0$.

We can use this to establish that the cone and wedge contours have the same intersection numbers with the thimble extending from the saddle point $z^{(0)}$ at the origin\footnote{Here we are being slightly sloppy. The saddle point at the origin is degenerate, and strictly speaking we ought to break this degeneracy---and any others---by introducing a small perturbation in the action before we can speak of a unique thimble decomposition. However, this does not change the results or any of the reasoning, so we have elided this step to keep the explanation brief and manageable.}. In the two-site case, consider the contour defined by the difference between the cone and wedge contours; we will show that this contour has intersection number $n_0 = 0$.

Using the notation of Section~\ref{sec:warmup}, the integration contour that gives the difference between the wedge and cone contours is
\begin{equation}
    \gamma_{\mathrm{wedge}} - \gamma_{\mathrm{cone}}
    = \gamma_3 \times (\gamma_1 - \gamma_4) + \gamma_4 \times (\gamma_2 - \gamma_3)
    \text.
\end{equation}
As-is, this contour does intersect the origin, and therefore contains a point on which the action is equal to that at the saddle point. However, the attentive reader may already observe that the contour intersects the origin \emph{twice}, with opposing orientations.

To make clear that the origin has no contribution, we can infinitesimally deform the contours $(\gamma_1 - \gamma_4)$ and $(\gamma_2 - \gamma_3)$ away from the origin. This increases the real part of the action at every point, and therefore results in a contour where the inequality $S(z) > S(z_0)$ is strict. With respect to this contour, the intersection number with the trivial saddle point must vanish: $n_0 = 0$.

\subsection{Semiclassics}
In the previous sections, we found numerically that the partition function for the analytically continued Hermitian theory differs from the partition function of the $\cal PT$-symmetric theory, but that this difference becomes exponentially small for high temperatures, cf. figure \ref{fig:compare}. In this section, we consider the high-temperature limit of the analytically continued theory by performing a semi-classical evaluation of the path integral. Note that at high temperature, the semi-classical evaluation is a good approximation because quantum fluctuations are highly suppressed.

To wit, when appropriately rescaling $\tau$ and $\phi$, the analytically continued partition function is given by
\be
\label{Zd116}
Z(\lambda=-g)=\int{\cal D}\phi\, e^{-S}\,,\quad S=\int_0^1 d\tau\left[\frac{\dot\phi^2}{2}-g \beta^3\phi^4\right]\,,
\ee
subject to periodic boundary conditions $\phi(0)=\phi(1)$. In the high temperature limit, we may attempt to evaluate this partition function by functional saddle-point method. Specifically, we have
\be
\phi(\tau)=\phi_{\rm cl}(\tau)+\phi^\prime(\tau)\,,\quad S=S^{(0)}+S^{(1)}+S^{(2)}+\ldots\,,
\ee
where $S^{(0)}=S[\phi=\phi_{\rm cl}]$ and 
\be
S^{(1)}=\int_0^1 d\tau \phi^\prime \left[-\ddot{\phi}_{\rm cl}-4 g \beta^3 \phi_{\rm cl}^3\right],
\ee
\be
S^{(2)}=\int_0^1 d\tau \frac{\phi^\prime}{2}\left[-\ddot{\phi}^{\prime}- 12 g \beta^3 \phi_{\rm cl}^2{\phi^\prime}\right]\\.
\ee
The saddle point condition of vanishing $S^{(1)}$ leads to the classical equations of motion
\be
    \ddot{\phi}_{\rm cl} = -4 g \beta^3 \phi_{\rm cl}^3.
\ee
The classical solution $\phi_{\rm cl}$ is given by 
\be
\label{phicl}
    \phi_{\rm cl}(\tau) = \frac{\Omega}{\sqrt{4 g \beta^3}}\cn\left(\Omega\tau+ B, \frac{1}{2}\right),
\ee
where $\cn$ denotes the Jacobi Elliptic cn function and $\Omega,B$ are two constants. We may recast the path integral in terms of these constants as follows. Writing
\be
Z=\int \frac{d\phi_i d\phi_f}{\sqrt{2\pi}} \delta(\phi_i-\phi_f) {\cal D}\phi^\prime e^{-S}\,,
\ee
where $\phi_i=\phi_{\rm cl}(0)$, $\phi_f=\phi_{\rm cl}(1)$, and we perform a change of variables $\phi_i,\phi_f\rightarrow \Omega,B$ such that
\be
Z=\int \frac{d\Omega dB}{\sqrt{2\pi}} \delta(\Omega-\Omega_n) \left|{\rm sn}\left(B,\frac{1}{2}\right) {\rm dn}\left(B,\frac{1}{2}\right)\right|\int {\cal D}\phi^\prime e^{-S}\,,
\ee
where ${\rm sn},{\rm dn}$ denote the Jacobi Elliptic sn, dn functions and $\Omega_n=4 n K\left(\frac{1}{2}\right)$ with $K(m)$ the complete elliptic integral of the first kind with modulus $m$. Here $\Omega_n$ with $n\in {\cal N}$ denotes periodic frequency of the Jacobi Elliptic functions that results from the periodicity requirement $\delta(\phi_i-\phi_f)$. Effectively, the integral over the constant $\Omega$ turns into a sum over $n$,
\be
Z=\sum_{n=0}^\infty\int \frac{dB}{\sqrt{2\pi}} \left| {\rm sn}\left(B,\frac{1}{2}\right) {\rm dn}\left(B,\frac{1}{2}\right)\right|\int {\cal D}\phi^\prime e^{-S}\,.
\ee
Restricting the classical solution (\ref{phicl}) to $\Omega=\Omega_n$ leads to the classical action
\be
S^{(0)}=\frac{\Omega_n^4}{48 g \beta^3}\simeq \frac{63.02}{g \beta^3}\times n^4\,.
\ee
Note that this correponds to a bounce contribution proportional to $e^{-3.98 / (\beta^3 \lambda)}$, consistent with the fit performed in Figure~\ref{fig:fit}.

The functional integration over the fluctuations $\phi^\prime$ can be calculated using the Gelfand-Yaglom method, cf.~\cite{Dunne_2008}. From $S^{(2)}$ above, the equations of motion for $\phi^\prime$ are
\be
\label{seco}
\ddot \phi^{\prime}=-12 g \beta^3 \phi_{\rm cl}^2 \phi^\prime\,,
\ee
with $\phi_{\rm cl}$ given by (\ref{phicl}) and Dirichlet boundary conditions $\phi^\prime(0)=\phi^\prime(1)=0$ because of $\phi=\phi_{\rm cl}+\phi^\prime$ and $\phi_{\rm cl}(0)=\phi(0)$, $\phi_{\rm cl}(1)=\phi(1)$. The Gelfand-Yaglom method implies
\be
\int {\cal D}\phi^\prime e^{-S^{(2)}}=\left[u(1)\right]^{-\frac{1}{2}}\,,
\ee
where $u(\tau)$ is a solution to (\ref{seco}) with different boundary conditions $u(0)=0,\dot u(0)=1$. The general solution to (\ref{seco}) can be found by the variation of the classical solution (\ref{phicl}), $u(\tau)=\delta \phi_{\rm cl}$ with respect to the parameters $\Omega,B$.  The solution fulfilling the boundary conditions can then be constructured straightforwardly, and one finds
\be
\label{det}
u(1)={\rm sn}^2\left(B,\frac{1}{2}\right) {\rm dn}^2\left(B,\frac{1}{2}\right)\,.
\ee
Putting everything together, we find in the semi-classical limit
\be
\label{Zcl}
Z(\lambda=-g)=\sum_{n=0}^\infty (-1)^n e^{-\frac{\Omega_n^4}{48 g \beta^3}}\left(\frac{2 K\left(\frac{1}{2}\right)}{\sqrt{2\pi}}+{\cal O}(g)\right)\,,
\ee
where we have taken the integral limits for $B$ to correspond to the points where ${\rm cn}\left(B,\frac{1}{2}\right)=\pm 1$. The origin of the factor $(-1)^n$ can be understood as follows: Regarding $-\partial_\tau^2-12 g \beta^3 \phi_{\rm cl}^2$ as a Schr\"odinger operator, we see that for $n=0$, the spectrum of the operator is real and positive, so the square root of the determinant is positive. For $n>0$, we can identify a zero-energy solution for the special case $B=0$ that fulfills the boundary conditions $\phi^\prime(0)=\phi^\prime(1)=0$ with wave-function $u(\tau)={\rm sn}\left(\Omega_n,\frac{1}{2}\right) {\rm dn}\left(\Omega_n \tau,\frac{1}{2}\right)$. For $n=1$, this wave-function has one node. It is well-known that the ground-state wavefunction for the Schr\"odinger equation has no nodes, so there must be exactly one energy eigenstate with $E<0$ for $n=1$ and $B=0$. If $B\neq 0$, the energy of the first excited state must also be negative, otherwise the determinant of the operator calculated in (\ref{det}) would have to be negative. As a result, we find that for $B\neq 0$ and $n=1$, there must be two negative eigenenergies, and and hence the sign of ${\rm det}^{-\frac{1}{2}}$ must be negative. For $n=2$, one can repeat this exercise, now noting that for $B=0$ has three nodes, and hence there must be four negative energy states for $B>0$, $n=2$. This generalizes to higher $n$, leading to the factor of $(-1)^n$ shown in (\ref{Zcl}).

We recognize (\ref{Zcl}) have the typical form expected for bounces, with $n=0$ the zero-bounce (perturbative) contribution, $n=1$ the one-bounce contribution, and $n>1$ multi-bounce contributions.

\subsection{N-component scalars in the large N limit}

Finally, let us consider quantum mechanics in $N$ dimensions, for which the Hermitian partition function reads
\be
\label{zd1def}
Z_{N,{\cal H}}^{d=1}=\int {\cal D}\vec{\phi} e^{-\int_0^\beta d\tau \left[\frac{1}{2}\dot{\vec\phi}^2+\frac{\lambda}{N}(\vec\phi^2)^2\right]}\,.
\ee
Using a Hubbard-Stratonovich transformation introducing the auxiliary field $\zeta$, this can be rewritten as in~\cite{Romatschke:2019rjk}, so that after performing the Gaussian integral over $\vec{\phi}$ one has
\be
\label{p111}
Z_{N,{\cal H}}^{d=1}=\int {\cal D}\zeta e^{-\int_0^\beta d\tau \frac{N \zeta^2}{4\lambda}-\frac{N}{2}{\rm tr}\ln \left[-\partial_\tau^2+2 i \zeta\right]}\,.
\ee
At large N, only the zero mode of the field $\zeta$ contributes; if in addition we limit our consideration to low temperatures, we have (cf. \cite{Romatschke:2019ybu})
\be
Z_{N\gg 1,{\cal H}}^{d=1}(\beta\rightarrow \infty)=\int d\zeta_0 e^{-N \beta\left(\frac{\zeta_0^2}{4\lambda}+\sqrt{\frac{i \zeta_0}{2}}\right)}\,.
\ee
At large $N$, the last integral may be calculated exactly using the saddle point method. There is only one saddle on the principal Riemann sheet, located at $i \zeta_0=2^{-\frac{1}{3}}\lambda^{\frac{2}{3}}$. One can identify the stable thimble connecting this saddle to the real line by the same technique that was used in appendix \ref{app}. Evaluating the action at the saddle, one thus has $\ln Z^{d=1}_{ N\gg 1,{\cal H}}(\beta \rightarrow 0)= -\beta E_0^{(N\gg 1)}$ where to leading order in large N, $E_0^{N\gg 1}=\frac{3 (2\lambda)^{\frac{1}{3}}}{8} N$. One can also calculate the contribution of order ${\cal O}(N^0)$ to $E_0$ as follows: expanding the partition function (\ref{p111}) to second order in fluctuations around the saddle: $\zeta=\zeta_0+\zeta^\prime(\tau)$ and performing a Fourier-transform on the fields $\zeta^\prime$, we obtain the fluctuation action in the small temperature limit as $S_2=\frac{1}{2}\int \frac{dk}{2\pi}\left|\zeta^\prime(k)\right|\left(\frac{1}{2\lambda}+\Pi(k)\right)$. Here $\Pi(k)=\frac{1}{2}\int \frac{dp}{2\pi}G(p)G(p+k)$, with $G^{-1}(p)=p^2+(2 \lambda)^{\frac{1}{3}}$ such that 
\be
\Pi(k)=\frac{1}{(2\lambda)^{\frac{1}{3}}(k^2+4 (2\lambda)^{\frac{2}{3}})}\,.
\ee
Performing the path integral over $\zeta^\prime$ leads to an expression for the spectral gap accurate to NLO in large N:
\be
\label{largeN}
E_0^{N\gg 1}=(2\lambda)^{\frac{1}{3}}\left(\frac{3 }{8} N+\frac{\sqrt{6}-2}{2}\right)+{\cal O}(N^{-1})\,.
\ee

A similar calculation can be performed for the 'wrong sign' partition function defined on the 'wedge' contour. Starting with the partition function 
\be
Z^{d=1}_{{\rm wedge} N}=\int_{\cal C} {\cal D}\vec{\phi} e^{-\int_0^\beta d\tau\left[
    \frac{1}{2}\dot{\vec{\phi}}^2-\frac{g}{N}\left(\vec{\phi}^2\right)^2\right]}\,,
\ee
with $\vec{\phi}$ a complex function of real-valued vectors $\vec{s}$ an obvious generalization to (\ref{Zd1wedge}) to N-components. Since $\vec{s}$ is a real-valued vector field, we introduce a Hubbart-Stratonovic transformation just as in the Hermitian theory case. Since the integral over $\vec{s}$ is again Gaussian, we find
\be
\label{ZPTd1}
Z^{d=1}_{{\rm wedge}, N}=\int {\cal D}\zeta \left[\frac{1}{2} e^{-\int_0^\infty d\tau \frac{N \zeta^2}{4 g} -\frac{N}{2}{\rm tr}\ln\left[-\partial_\tau^2-2 \zeta\right]}+\zeta\rightarrow -\zeta\right]\,,
\ee
which is still exact for all N. In the large N limit, we can again use the fact that the partition function can be evaluated from the saddle points of the action, which is the Fourier zero mode $\zeta(\tau)=\zeta_0$. The calculation then proceeds exactly analogous to the Hermitian case, even though the saddle point locations $\zeta_0$ are complex. One finds
\ba
\label{wedgeE}
E_{{\rm wedge},0}^{N\gg1}(g)&=& \frac{(2 g)^{\frac{1}{3}}}{2} \left(\frac{3}{8} N +\frac{\sqrt{6}-2}{2}\right) \\
&&-\frac{1}{\beta}\ln\left[2 \cos \left(\frac{\sqrt{3}N (2 g \beta^3)^{\frac{1}{3}}}{16}\right)\right]+ {\cal O}(N^{-1})\,.\nonumber
\ea

Comparing \ref{wedgeE} and (\ref{largeN}), we find that in the zero temperature limit to leading and NLO order in large N,
\be
\ln Z_{\rm wedge}(g)={\rm Re}\ln Z_{\cal H} (\lambda=-g)\,.
\ee

\section{Discussion}
\label{sec:disc}

In this work, we have examined the relation between interacting quantum theories with quartic interaction. Specifically, we have studied if and how analytically continuing the Hermitian theory to negative coupling can be related to the ${\cal PT}$-symmetric theory.

Based on our detailed calculations performed in $d=0$ and $d=1$, our findings are as follows:
\begin{itemize}
\item
We showed that a path-integral formulation on a complex field contour (the 'wedge') for the 'wrong sign' Hermitian theory has the property that its partition function equals the real part of the analytically continued Hermitian theory (\ref{conj}).
\item
We found that this complex integration contour (the `wedge') is different from---and yields a different integral than---the complex integration contour used to define the ${\cal PT}$-symmetric theory (the `cone').
\item
We found that the difference in integration contours corresponds to a non-perturbative contribution to the partition function (the `bounce'). Evaluating the leading bounce contribution analytically using semiclassics, we find excellent numerical agreement with the difference between the partition functions defined on the two contours.
\item
We provided evidence from high-precision numerical calculations that the path integrals defined on the `wedge' and `cone', respectively, correspond to the partition function calculated from the known spectrum of the Hamiltonian for the analytically continued Hermitian and ${\cal PT}$-symmetric theories.
\item
We found that because the bounce contribution becomes exponentially suppressed at high temperature (equivalently, weak coupling), the partition functions defined on the two integration contours are exponentially close in that limit.
\item
We found that in the limit of a large number of fields, the difference between the relations (\ref{ABS}) and (\ref{conj}) becomes large-$N$ suppressed
\end{itemize}

Based on these findings, we offer the following interpretations concerning the relation between analytically continued Hermitian and $\cal PT$-symmetric field theory:
\begin{itemize}
\item
The ABS conjecture (\ref{ABS}) is likely incorrect. In all cases we studied, (\ref{ABS}) was violated, for reasons we have outlined in this work.
\item
The relation (\ref{conj}) holds to very good approximation at high temperatures. This is because the non-perturbative corrections to the left-hand side of (\ref{conj}) are  exponentially suppressed at high temperature.
\item
The analytically continued Hermitian partition function does have a consistent formulation as a path integral on a complex integration contour, it is just not the ${\cal PT}$-symmetric integration contour. This `wedge' contour gives the exact analytic continuation of the Hermitian theory for all temperatures and all number of field components.
\item
In the large volume (zero temperature) limit, we expect the relation
\be
\label{mainrel}
\ln Z_{\rm wedge}(g)={\rm Re}\ln Z_{\cal H}(\lambda=-g)\,,
\ee
which we proved for $d=0,1$, to leading and next-to-leading order in $\frac 1 N$ to generalize to arbitrary dimension $d$.
\end{itemize}

While the original ABS conjecture does not seem to hold, we believe that the existence of the relation (\ref{mainrel}) puts the analytic continuation of `wrong sign' field theories such as those discussed in Refs.~\cite{Romatschke:2022jqg, Romatschke:2022llf,Grable:2023paf} on firm footing.

\begin{acknowledgements}
We would like to thank Wen-Yuan Ai, Carl Bender, Seth Grable, Sarben Sarkar and Max Weiner for helpful discussions. This work was supported by the Department of Energy, DOE award DE-SC0017905. 
\end{acknowledgements}

\appendix
\section{One-site calculations for N-component scalars}\label{app}

In this section, we provide some calculational details for the case of N-component scalars in d=0 discussed in section \ref{sec:warmup} in the main text. To start, note that (\ref{d0hermitz}) can be calculated using spherical coordinates in N dimensions. This leads to
\be
\label{d0exact}
Z_{N}^{d=0}=\frac{2^{1-\frac{N}{2}}}{\Gamma\left(\frac{N}{2}\right)}\int_0^\infty dr r^{N-1} e^{-\frac{\lambda}{N}r^4}=\left(\frac{4\lambda}{N}\right)^{-\frac{N}{4}}\frac{\Gamma\left(\frac{N}{4}\right)}{2\Gamma\left(\frac{N}{2}\right)}\,.
\ee
In particular, for large $N\gg 1$, the asymptotic expansion for the $\Gamma$-function then leads to
\be
\label{Zd0ex}
\ln Z^{d=0}_{N\gg 1}=-\frac{N}{4}\ln \frac{4\lambda}{e^{1}}-\frac{\ln 2}{2}+{\cal O}(N^{-1})\,.
\ee
This large N behavior may also be obtained directly using the method of steepest descent. To this end, rewrite
\be
Z^{d=0}_{N}=\sqrt{\frac{N}{4 \lambda \pi}}\int \frac{d\vec{\phi}}{(2\pi)^{\frac{N}{2}}} \int_{-\infty}^\infty d\zeta e^{- i \zeta \vec{\phi}^2-\frac{\zeta^2 N}{4\lambda}}\,.
\ee
Now the integral over $\vec{\phi}$ is Gaussian and can be done exactly to give
\be
Z^{d=0}{N}=\sqrt{\frac{N}{4 \lambda \pi}}2^{-\frac{N}{2}}\int_{-\infty}^\infty d\zeta e^{-\frac{\zeta^2 N}{4\lambda}-\frac{N}{2} \ln (i\zeta)}\,.
\ee
For $N\gg 1$, the integral can be evaluated exactly using the method of steepest descent. The saddle point condition then is
\be
\frac{\zeta}{2\lambda}+\frac{1}{2\zeta}=0\,,
\ee
which is solved by $\zeta=\pm i \sqrt{\lambda}$. To find out which saddle contributes to the path integral, we consider a path parametrized by $s$ in the complex $\zeta$-plane, so that $\zeta(s)=a(s)+i b(s)$ with real $a(s),b(s)$. The special path we are interested in is called a Lefschetz thimble, and it is defined through the solution of the flow equations (\ref{eq:hgf})
where here $S=\frac{\zeta^2}{4\lambda}+\frac{1}{2}\ln( i\zeta)$. The thimbles have the special property that the imaginary part of the action is constant along the thimble, which can easily be seen from noting that
\be
\frac{d S}{ds}=\frac{\partial S}{\partial \zeta}\frac{d \zeta}{d s}=\left|\frac{\partial S}{\partial \zeta}\right|^2\,.
\ee
For the saddle located at $\zeta=-i \sqrt{\lambda}$, $S$ is real, which can be used to find the corresponding thimbles passing through this saddle without solving (\ref{eq:hgf}). Specifically, one finds that one thimble is given by
\be
a(s)=\sqrt{\lambda} x(s) \sqrt{\frac{{\atan} x(s)}{x(s)}}\,,
b(s)=-\sqrt{\lambda}\sqrt{\frac{{\atan} x(s)}{x(s)}}\,,
\ee
with $x(s)=\pm e^s$ on the right and left part of the thimble. There is also an unstable thimble given by $a(s)=0,b=-s$ with $s>0$, but this thimble does not connect to the real line at $\zeta=\pm \infty$, so it is dismissed. For the second saddle at $\zeta=+i \sqrt{\lambda}$, one finds that in the complex $\zeta$ plane the branch cut of the logarithm implies that these are actually multiple saddles on different Riemann sheets. Not surprisingly, there are no thimbles that connect to the real line on the principal Riemann sheet and go through these saddles, so there is no contribution from the saddle at $\zeta=\pm \infty$ to the path integral.

Using the stable thimble through the saddle $\zeta=-i \sqrt{\lambda}$, expanding $S[\zeta]$ to quadratic order and doing the Gaussian integral then gives
\be
\ln Z^{d=0}_{N\gg 1}=-\frac{N}{4} \ln \frac{4\lambda}{e^{1}}-\frac{1}{2}\ln 2  +{\cal O}(N^{-1})\,,
\ee
matching the large N limit of the exact result (\ref{Zd0ex}).

We close this section by giving detailed results for the d=0 partition function for N-components on the 'wedge' contour. Explicitly, in the case of N=2, we have for this choice of contour
\be
Z_{{\rm wedge},N=2}^{d=0}=\int_0^\infty \frac{ds dt}{2\pi} e^{-\frac{g}{2}(s^2+t^2)}\left(2 e^{-\frac{i \pi}{2}}+2 e^{\frac{i \pi}{2}}\right)=0\,.
\ee
Comparing this result to the Hermitian O(N) result (\ref{d0exact}), one finds that the path integral defined on the 'wedge' exactly matches the real part of the analytically continued Hermitian result.

For three or more components, one proceeds in a similar fashion to find
\be
Z_{{\rm wedge},N}^{d=0}=\frac{2^{1-\frac{N}{2}}}{\Gamma\left(\frac{N}{2}\right)}\int_0^\infty dr r^{N-1} e^{-\frac{g}{N}r^4}\frac{\left(e^{-\frac{N i \pi}{4}}+ e^{\frac{N i \pi}{4}}\right)}{2}\,,
\ee
which proves 
\be
\label{con1}
Z_{{\rm wedge},N}^{d=0}={\rm Re}\left\{Z_{N}^{d=0}(\lambda=-g+i 0^+)\right\}\,,
\ee
for all $N\in \mathbb{N}$.
At large $N\gg 1$, one notes that
\be
\ln Z_{{\rm wedge},N\gg 1}^{d=0} =-\frac{N}{4} \ln \frac{4g}{e^{1}}-\frac{\ln 2}{2} +\ln \cos \frac{N \pi}{4} + {\cal O}(N^{-1})\,.
\ee
Since the logarithm of the cosine is not proportional to $N$, in addition to (\ref{con1}), at large N the d=0 theory fulfills the additional relation
\be
\label{conjecture}
\ln Z_{{\rm wedge},N\gg 1}^{d=0} ={\rm Re}\ln Z^{d=0}_{N\gg 1}(\lambda =-g+i 0^+)+{\cal O}(N^{0})\,.
\ee
Thus at large N, and large N only, the Hermitian and wedge-contour parametrized partition functions for the d=0 case are related through the original conjecture (\ref{conj}).

\bibliography{PT}
\end{document}